\documentclass[proof]{pasj01}
\draft 
\Received{$\langle$reception date$\rangle$}
\Accepted{$\langle$acception date$\rangle$}
\Published{$\langle$publication date$\rangle$}

\begin{document}

\title{Gaia DR 2 data and the evolutionary status of eight high velocity hot post-AGB candidates}

\author{Mudumba \textsc{Parthasarathy}\altaffilmark{1}\altaffilmark{2}}
\altaffiltext{1}{Indian Institute of Astrophysics, II Block, Koramangala, Bangalore 560 034, INDIA}
\altaffiltext{2}{National Astronomical Observatory, 
 2-21-1 Osawa, Mitaka, Tokyo 181-8588, Japan }
\email{m-partha@hotmail.com}

\author{Tadafumi \textsc{Matsuno}\altaffilmark{3}}
\altaffiltext{3}{Department of Astronomical Science, School of Physical Sciences, The Graduate University of Advanced Studies (SOKENDAI), 2-21-1 Osawa, Mitaka,
Tokyo 181-8588, Japan}
\email{matsuno@astro.rug.nl}

\author{Wako \textsc{Aoki}\altaffilmark{2}\altaffilmark{3}}
\email{aoki.wako@nao.ac.jp}


\KeyWords{stars:evolution --- stars:AGB and post-AGB --- stars:high-velocity --- stars:distances}

\maketitle

\begin{abstract}
From {\it Gaia} DR 2 data of eight high velocity hot post-AGB candidates LS
3593, LSE 148, LS 5107, HD 172324, HD 214539, LS IV -12 111,
LS III +52 24, and LS 3099, we found that six of them
have accurate parallaxes which made it possible to derive their
distances, absolute visual magnitudes ($M_{V}$) and luminosity 
($\log L/L_{\odot}$). Except LS 5107 all the remaining seven stars have
accurate effective temperature ($T_{\rm eff}$) in the
literature. Some of these stars are metal-poor and
some of them do not have circumstellar
dust shells. In the past the distances of some stars were
estimated to be 6~kpc which we find it to be incorrect. The accurate
{\it Gaia} DR2 parallaxes show that they are relatively nearby
post-AGB stars.
When compared with post-AGB evolutionary tracks we find their initial
masses in the range of 1~M$_{\odot}$ to 2~M$_{\odot}$. We find the
luminosity of LSE 148 to be significantly lower than that of post-AGB
stars, suggesting that this is a post-horizontal branch star or
post-early-AGB star. LS 3593 and LS 5107 are new 
 high velocity hot post-AGB stars from {\it Gaia} DR2.  

\end{abstract}

\section{Introduction}
Hot post-AGB stars are transition objects evolving towards the early
stages of planetary nebulae (PNe) (\cite{kwok93}; \cite{vanwinckel03}; \cite{parthasarathy06}; \cite{parthasarathy93,
  parthasarathy95}, and references therein). Some of the hot post-AGB
stars are at high galactic latitudes and a few are high velocity stars, 
e.g., LS III +52 24 (IRAS 22023+5249: $V_{\rm r} = -148$ km~s$^{-1}$;
\cite{sarkar12}), LS 5112 (IRAS 18379-1707: $V_{\rm r}=-124$
km~s$^{-1}$; \cite{ikonnikova19}). These stars could be highly evolved low-mass metal-poor stars in the halo population of the Milky Way. Determination of evolutionary status and frequency of these objects is useful for better understanding of both low-mass star evolution and the Milky Way halo structure.

Hence, these objects prompted us to look
for more such stars. However the radial velocities and distances of
most of the hot post-AGB candidates were not available. With the
advent of {\it Gaia} DR 2 data \citep{gaia18, lindegren18} one can look
for more such stars.

We considered the hot post-AGB candidates given in the following
papers: \citet{mello12, venn98, kendall94, przybylski69, klochkova18,
  sarkar05, sarkar12}. From the {\it Gaia} DR 2 data we found LS 3593
(SAO 243754), LSE 148 (HD 177566), LS 5107 (IRAS 18365-1353), HD
172324 (V 534 Lyr), HD 214539, and LS 3099 to be high velocity stars
with accurate {\it Gaia} DR 2 parallaxes (see Tables 1 and 2).  LS IV
-12 111 and LS III +52 24 do not have accurate parallaxes. We also
looked at the proper motion data of these eight stars in the {\it
  Gaia} DR 2.

In this paper we analyze the {\it Gaia} DR 2 data of the above mentioned eight
stars and based on it we discuss their evolutionary status.

\section{Data and analysis}

The Galactic longitude and latitude, parallaxes, radial velocities, $G$ ({\it Gaia} $G$ band), $V$,
$B-V$ and spectral types of all the eight stars mentioned above are given
in Table 1. The data are taken from SIMBAD and {\it Gaia} DR2.


Using the {\it Gaia} DR 2 parallaxes the distances are derived (Table
2). The $E(B-V)$ values (Table 2) are obtained from \citet{schlegel98}
for high galactic latitude stars (LSE 148, HD172324,
HD214539). The interstellar reddening of these three stars is quite small. \citet{chen19} and \citet{green18} are used for LS3593 and
LS5107, respectively. Extinction coefficients
are converted using values provided in \citet{green18},
\citet{schlafly11}, and mean values in \citet{casagrande18}.  Table 2
lists $E(B-V)$ obtained from dust map as well as $E(B-V)$ values
calculated from the observed $(B-V)$ and intrinsic $(B-V)$ obtained
from their spectral types ($E(B-V)_{\rm sp}$).  The
$E(B-V)_{\rm sp}$ values of three IRAS sources are 0.09 -- 0.32 mag 
larger than the $(E-V)$ estimated from the dust map. This suggests that they are affected by
circumstellar reddening, in addition to the interstellar reddening.


The calculated absolute visual magnitudes $M_{V}$ are given in Table
2.  For $M_{V}$ values we applied bolometric corrections as the
$T_{\rm eff}$ values and spectral types of all the stars are well
determined. The bolometric corrections were taken from Allen's
Astrophysical Quantities (4th edition; \cite{cox00}). The values of the bolometric corrections of \citet{flower96} are 0.1-0.25~magnitude larger than those of the Allen's Astrophysical Quantities, resulting in difference in $\log L/L_{\odot}$ by less than 0.1~dex. This does not affect the discussion on the nature of our high velocity post-AGB stars. The derived
bolometric magnitudes $M_{\rm bol}$ and $\log L/L_{\odot}$ values are
given in Table 2. Notes on all the eight stars are given below.

Stellar parameters, i.e. $T_{\rm eff}$, $\log g$, and [Fe/H] are given
in Table 2. Details of these parameters and references will be
provided for individual objects below.

Kinematics are calculated using {\it Gaia} parallax and proper motion
measurements. We adopt 8.2 kpc as the distance between the Sun and the
Galactic center \citep{mcmillan17}, 0.025 kpc as the vertical offset
of the Sun \citep{juric08}. Solar motion is adopted from
\citet{schonrich10} for the radial and vertical velocities (11.1
km~s$^{-1}$ and 7.25 km~s$^{-1}$, respectively) and is calculated as
247.97 km~s$^{-1}$ using the proper motion measurements by
\citet{reid04}. The results are  presented in Table 3.

From the radial velocity measurements by previous studies, no clear signature of binarity has been found for these stars. The constraint is still not very strong due to the limitation of the number of spectroscopic observations. We note that the sample selection of high velocity post-AGB stars would not be affected if they belong to low-mass star binaries because the radial velocity variations expected for low-mass star binaries are not as large as the radial velocity of the stars studied in this paper.

\subsection{Notes on the eight stars}

\begin{itemize}

\item{LS 3593}

\citet{venn98} derived $T_{\rm eff} = 9300$~K, $\log g = 1.7$. They have
derived chemical composition of this star. They found it to be
metal-poor ([Fe/H]$ = -2.0$).  Oxygen and nitrogen seem to be slightly
overabundant.  They do not find any emission lines in the
spectrum. They found it to be a very slow rotator. Its galactic
latitude is low but is found to be a high velocity star.

\item{LSE 148}

It is a high galactic latitude and high velocity star (Table
1). \citet{kendall94} derived $T_{\rm eff} = 30,600$~K and $\log g =
3.5$. They find it to be metal-poor. Carbon is underabundant similar
to that observed in high galactic latitude hot post-AGB stars. They
conclude that it is a hot post-AGB star of core mass
0.55M$_{\odot}$. They find it to be a very slow rotator.  No emission
lines are present in the spectrum.  \citet{mello12} also find it to be
very metal-poor.  They find weak emission in the cores of Balmer
lines. \citet{mello12} derived $T_{\rm eff}$ = 30,900~K, $\log g =3.8$
which agrees well with the results obtained by \citet{kendall94}.
\citet{mello12} find no Fe lines in the spectrum of LSE 148. They
suggested that it may be a post-horizontal branch star or a post-early
AGB star. This is supported by the present work with much robust
estimate of the luminosity (see Sect. 3).

\item{LS 5107 (IRAS 18365-1353)}

There is no detailed study of the spectrum of this
star. \citet{venn98} find that H$\alpha$ line to be a P-Cygni profile
and find Fe II emission line at 6515~{\AA}. LS 5107 is found to be an
IRAS source with far-IR colors similar to that of post-AGB stars and
planetary nebulae. The $T_{\rm eff}$ given in Table 2 is estimated
from its spectral type.  We conclude that LS 5107 is a high velocity
hot post-AGB star.

\item{HD 172324 (V534 Lyr)}

  It is a high galactic latitude and high velocity star (Table 1). It
  is a small amplitude light and radial velocity variable. Recently,
  \citet{klochkova18} made detailed analysis of the high resolution
  spectra of this star and derived its chemical composition. They find
  $T_{\rm eff}$ = 10,000K, $\log g = 2.5$. They find it to be
  metal-poor and overabundant in nitrogen. The H$\alpha$ line shows
  variable P-Cygni profile.  They conclude that it is a population-II
  pulsating star near the horizontal branch. \citet{klochkova18}
  derived a distance of 6 kpc, whereas \citet{bonsack56} estimated a
  distance of 5.7 kpc. The {\it Gaia} DR 2 data (Tables 1 and 2),
  however, clearly shows that their distance estimate is wrong. 
    The luminosity estimated from this distance is well higher than
    the value expected for the horizontal branch stars. We note that
    the $\log g$ value derived by spectroscopic analysis by
    \citet{klochkova18} is also lower than the typical values of
    horizontal branch stars.

\item{HD 214539}

It is a high galactic latitude and very high velocity star (Table
1). \citet{przybylski69} made a detailed analysis of the spectrum of
this star. He finds it to be a metal-poor ([Fe/H]$= -1.2$) and low
gravity star.  It is considered at that time the brightest known blue
halo star. \citet{kodaira84} derived $T_{\rm eff}$ = 9800~K and $\log
g=1.6$.

\item{LS 3099 (IRAS 13266-5551)}

\citet{mello12} and \citet{sarkar05} analyzed the high resolution
spectrum of this B1 Iae post-AGB star.  They find $T_{\rm
  eff}=20,200$~K, $\log g = 2.38$ and mildly carbon-poor and
metal-poor \citet{mello12}.  The radial velocity is found to be
65.31$\pm$0.34 km~s$^{-1}$ \citep{sarkar05}.

\item{LS IV -12 111 (IRAS 19590-1249)}

It is a high galactic latitude ($b = -21.26$ degrees) and high
velocity (86.9 km~s$^{-1}$) hot (B1Iae) post-AGB star. \citet{ryans03}
and \citet{mello12} analyzed high resolution spectrum of this
star. They derive $T_{\rm eff}=20,500$~K and $\log g = 2.5$
\citep{ryans03}. The values derived by \citet{mello12} are in
agreement with the results of \citet{ryans03}. They find it to be
carbon-poor by $\sim 0.4$ dex similar to that found in high galactic
latitude OB post-AGB stars, indicating that the star left the AGB
before the third dredge-up. The parallax of this object in {\it
  Gaia} DR2 is not sufficiently accurate.

\item{LS III +52 24 (IRAS 22023+5249)}

\citet{sarkar12} analyzed high resolution spectrum of this hot (B1 I)
post-AGB star. They find the radial velocity to be $-148.31\pm
0.60$ km~s$^{-1}$. They derive $T_{\rm eff} = 24,000 \pm 1000$~K,
$\log g = 3.0 \pm 0.5$ and $-0.12$ dex metal-poor. The spectrum shows
nebular emission lines indicating the presence of a low excitation
planetary nebula. \citet{arkhipova13} find small amplitude light
variations similar to that found in other post-AGB stars.
The parallax of this object in {\it Gaia} DR2 is not sufficiently accurate.

\end{itemize}

\section{Discussion}\label{sec:discussion}

We show in Figure 1 the location of six high velocity stars in the HR
diagram. LS IV -12 111 and LS III +52 24 parallaxes are not determined
with accuracy of three sigma, hence, they are not shown in Figure 1.
{Four of the six stars have similar $T_{\rm eff}$ and $L$ along the
  evolutionary tracks of low-mass post-AGB stars}.

Kinematics show that all stars except LS 5107 do not follow the
Galactic rotation. LS 5107, for which there is no estimate of
metallicity, could belong to the thin disk, according to the
kinematics information. LSE 148 and HD 214539 show large velocities
relative to the Sun, and hence they would belong to the Galactic halo,
indicating that their initial masses are low. The other three, LS 3593, 
HD172324 and LS 3099 show modest relative velocities compared to the
Sun. Although they could belong to the Galactic disk, it is unlikely
that their ages are young given the age-velocity dispersion relation
among Milky Way disk stars (e.g., \cite{casagrande11}).  These
  kinematics features are consistent with the results from the
  positions in the $T_{\rm eff}$-$\log L$ plane that these stars are
  low-mass halo post-AGB objects.

LSE 148 has luminosity that is even lower than that of low-mass post-AGB
stars. This suggests that this is a post-horizontal branch star or
post-early-AGB star. 

LS 3099 has luminosity explained by post-AGB stars with about
2.0M$_{\odot}$ and is not metal-poor. This star may be a thick disk
star with relatively high metallicity.

Although LS IV-12 111 and LS III +52 24 do not have reliable parallax
measurements, we can place lower limits on their tangential
velocities. The $2\sigma$ lower limits are 108 $\mathrm{(km\,s^{-1})}$
and 118 $\mathrm{(km\,s^{-1})}$. These results, together with their
large radial velocities, support that these stars are high-velocity
stars.

\subsection{Frequency of metal-poor post-AGB stars and timescale of evolution}

We find four metal-poor post-AGB stars with low core masses ($\sim
0.55$M$_{\odot}$, which corresponds to initial masses of
1.0M$_{\odot}$ or smaller). They have $T_{\rm eff} \sim 10000$~K.
  Taking account of the incompleteness of the sample of this study,
  this can be used to estimate the lower limit of the frequency of
  such objects in the Galaxy. The distances of these objects are
about 2~kpc or less, which became available by the {\it Gaia} DR2 for
the first time. 
The local stellar density of the halo structure is estimated to
  be $3-15\times 10^{-5}$~M$_{\odot}$~pc$^{-3}$ (\cite{deason19} and
  references therein). Adopting the recent estimate, $7 \times
  10^{-5}~$M$_{\odot}$~pc$^{-3}$, by \citet{deason19}, the stellar halo
  mass within 2~kpc is $1.7\times 10^{6}~$M$_{\odot}$. Hence, the
  frequency of metal-poor post-AGB stars is the order of 10$^{-6}$. If
  higher luminosity (larger distance) was applied as estimated by
  previous studies for some of our post-AGB stars, the frequency could be much
  lower.
Taking account of the lifetime of such low-mass stars is $\sim 10^{10}$ years, the time-scale of the
evolution in this phase is estimated to be 10$^{4}$ years. This
roughly agrees with the prediction of the recent model calculation of
\citet{miller16} shown in their Figure 8 for metal-poor low-mass
stars.

Since the timescale of the evolution of low-mass metal-poor objects is
even longer from 10000~K to the maximum effective temperature ($\sim
100,000~$K), hotter high velocity post-AGB stars with low-metallicity
could exist in the Milky Way halo that are yet identified. Further
searches for such stars will provide useful constraint on the
frequency of low-mass metal-poor post-AGB stars.

\begin{ack}

MP was supported by the NAOJ Visiting Fellow Program of the Research
Coordination Committee, National Astronomical Observatory of Japan
(NAOJ), National Institutes of Natural Sciences(NINS).
\end{ack}

\clearpage


\begin{table}
\tbl{Objects
\label{tab:obj}}{
\begin{tabular}{lrrrrrrrll}
\hline\noalign{\vskip 3pt}
Object & $l$ & $b$ & Parallax($p$) & $V_{\rm r}$ & $G$ & $V$ & $B-V$ & spectral type & \\
       & (degree) & (degree) & (mas) & (km~s$^{-1}$)  & (mag) & (mag) & (mag) & & Note\\
\hline\noalign{\vskip3pt} 
LS 3593   & 330.6439866 & -3.672106351 & 0.410$\pm$0.047 & $110.12\pm$1.94 & 9.44 & 9.53 & 0.15 & B8Iab & \\
LSE 148   & 355.54719 & -20.42079462 & 0.697$\pm$0.106 & $-134.00\pm$4.30 & 10.15 & 10.17 & $-0.20$ & B6Ib & \\
LS 5107   & 19.24285127 & -3.663147838 & 0.624$\pm$0.052 & $-76.65\pm$8.30 & 9.52 & 9.89 & 0.86 & B9Iae & V452 Sct\\
          &&&&&&&&& IRAS source \\
HD 172324 & 66.18384285 & 18.58124629 & 0.456$\pm$0.042 & $-117.10\pm$2.00 & 8.14 & 8.16 & 0.01 & A0Iabe & V534 Lyr\\
HD 214539 & 319.7541249 & -44.93459676 & 0.719$\pm$0.041 & $333.00\pm$2.80 & 7.18 & 7.25 & 0.00 & B9Ib/A0Iab & \\
LS 3099      & 308.3016934 & 6.35580537 & 0.377$\pm$0.050 & $65.31\pm$0.34 &10.70 & 10.68 & 0.39 & B1Iae & IRAS source \\
LS IV-12 111 & 29.17960346 & -21.2637799 & 0.129$\pm$0.055: & $86.9\pm$1.7  & 11.32 & 11.4 & $-0.1$ & B1Iae & IRAS source \\
LS III +52 24 & 99.30346632 & -1.955656239 &0.080$\pm$0.052: & $-148.31\pm$0.60 & 12.36 & 12.51 & 0.72 & B1I & IRAS source \\
\hline\noalign{\vskip 3pt} 
\end{tabular}
}
\end{table}

\begin{table}
\tbl{Distance and stellar parameters
\label{tab:obj2}}{
\begin{tabular}{lrrrrrrrrrr}
\hline\noalign{\vskip 3pt}
Object & Distance($D$)  & $E(B-V)$ & $E(B-V)_{\rm sp}$ & $M_{V}$ & $M_{g}$ & $\log L/L_{\odot}$ &  $T_{\rm eff}$ & $\log g$ & [Fe/H] & references\\
       & (pc) & (mag)  & (mag) & (mag) & (mag) & &  (K) & & & \\
\hline\noalign{\vskip3pt} 
LS 3593   & 2436$\pm$281 & 0.21 & 0.18 & -2.97 & -3.07$\pm$0.26 & 3.35$\pm$0.11 &  9300 & 1.7 & $-2.0$ & 1 \\
LSE 148   & 1436$\pm$219 & 0.10 & 0.07 & -0.08 & -0.91$\pm$0.34 & 2.59$\pm$0.14 & 30900 & 3.8 & $<-2.0$ & 2 \\
LS 5107   & 1602$\pm$135 & 0.56 & 0.88 & -2.64 & -3.04$\pm$0.19 & 3.17$\pm$0.09 & 10500 & 2.0 & ... & \\
HD 172324 & 2194$\pm$202 & 0.06 & 0.02 & -3.71 & -3.73$\pm$0.21 & 3.54$\pm$0.09 & 10000 & 2.5 & $-0.3$ & 3 \\
HD 214539 & 1392$\pm$80 & 0.03 & 0.02 & -3.55 & -3.62$\pm$0.14 & 3.53$\pm$0.07 &  9800 & 1.6 & $<-1.5$ & 4,5\\
LS 3099   &  2650$\pm$351 & 0.33 & 0.51 & $-3.12$  & ... & 3.67$\pm$0.12& 20200 & 2.38 & &  \\
LS IV-12 111 & $>4180$  & 0.19 & 0.28 & $<-2.32$ & $<-2.32$ & $>3.63$ & 20500 & 2.5  & &  \\
LS III 52-24 & $>5398$  & 0.67 & 0.64 & $<-2.95$ & $<-3.14$ & $>3.92$ & 24000 & 3.0  & & \\
\hline\noalign{\vskip 3pt} 
\end{tabular}
}
  \begin{tabnote}
References -- 1:\citet{venn98}; 2: \cite{mello12}; 3: \citet{klochkova18}; 4: \citet{kodaira84}; 5:\citet{przybylski69} 
  \end{tabnote}
\end{table}

\begin{table}
\tbl{Kinamatics data
\label{tab:kinematics}}{
\begin{tabular}{lrrrrrrrrrr}
\hline\noalign{\vskip 3pt}
Object & $\mu_{\alpha}$ & $\sigma(\mu_{\alpha})$ & $\mu_{\delta}$ & $\sigma(\mu_{\delta})$ & $V_{R}$ & $\sigma (V_{R})$ & $V_{\phi}$ & $\sigma(V_{\phi})$ & $V_{Z}$ & $\sigma(V_{Z}$) \\
       & (mas) & (mas) & (mas) & (mas) & (km~s$^{-1}$) & (km~s$^{-1}$) & (km~s$^{-1}$) &(km~s$^{-1}$) &(km~s$^{-1}$) &(km~s$^{-1}$) \\
\hline\noalign{\vskip3pt} 
LS 3593   & $-4.252$ & 0.066 & -11.621 & 0.053 & $-56.85$ & 7.44 & 78.13 & 11.73 & $-56.17$ & 6.27 \\
LSE 148   & $-2.030$ & 0.167 & -42.817 & 0.157 & $163.10$ & 7.89 & $-8.18$ & 34.46 & $-17.97$ & 10.04 \\
LS 5107   & $-4.487$ & 0.082 &  -6.045 & 0.074 & $-83.46$ & 7.52 & 228.88 & 4.71 & 11.15 & 0.83 \\
HD 172324 & $-3.183$ & 0.079 &  -0.969 & 0.080 & $ 68.09$ & 3.74 & 144.65 & 2.31 & $-61.38$ & 3.23 \\
HD 214539 & $-25.891$& 0.057 & -26.177 & 0.065 & $-320.38$ & 7.23 & $-31.02$ & 8.53 & $-63.93$ & 9.96 \\
LS 3099      & $-5.69$ & 0.074 & $-2.744$ & 0.079 &$-39.7$  & 2.0  & 146.7  & 5.1  & $-6.9$  & 3.2  \\
LS IV-12 111 & $-3.596$ & 0.094 & $-4.096$ & 0.061 & & & & & &  \\
LS III 52-24 & $-3.856$ & 0.100 & $-2.566$ & 0.070 & & & & & & \\
\hline\noalign{\vskip 3pt} 
\end{tabular}
}
\end{table}

\begin{figure}
 \begin{center}
  \includegraphics[width=10cm]{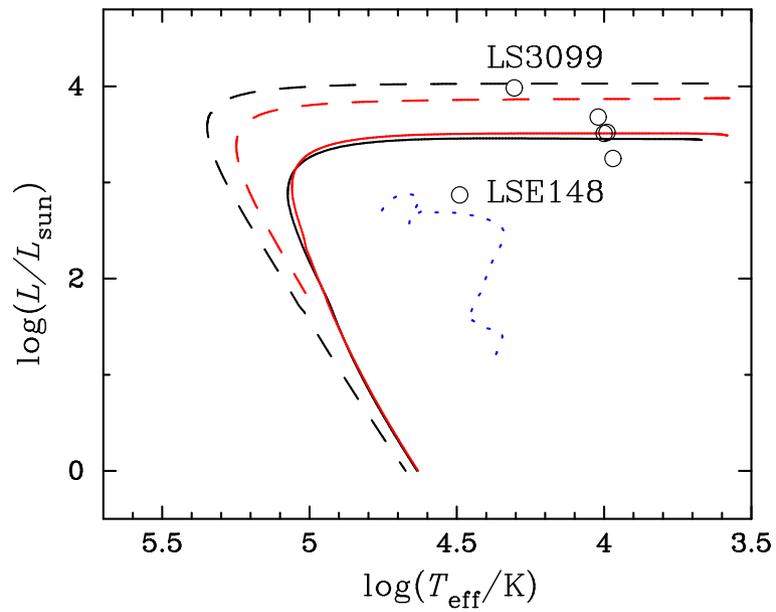}
 \end{center}
 \caption{Evolutionary tracks of post-AGB phases taken from
   \citet{miller16} for initial masses of 1.0~M$_{\odot}$ (solid
   lines) and 2.0~M$_{\odot}$ (dotted lines) with $Z=0.02$ (red) and
   $Z=0.001$ (black). The evolutionary track of post-horizontal branch
   star for the core mass of 0.52~M$_{\odot}$ with [Fe/H]$=-1.48$
   taken from \citet{dorman93} is shown by dotted (blue) line. The six
   object studied in the present work are shown by open
   circles.}\label{fig:track}
\end{figure}


\end{document}